\documentclass[showpacs,prl,floatfix,amsmath,amsfonts,superscriptaddress,twocolumn]{revtex4}
\usepackage{graphicx,amsmath,amsfonts}
\usepackage[dvips]{changebar}
\usepackage[usenames]{color}
\usepackage[]{hyperref}
\newcommand{\pd}[2]{\frac{\partial #1}{\partial #2}}

\newcommand{\eps}{\varepsilon}
\newcommand{\ud}{\mathrm{d}}

\begin{document}
\title{The piston dispersive shock wave problem}
\date{\today}
\author{M. A. \surname{Hoefer}}
\email{hoefer@boulder.nist.gov}
\thanks{Contribution of the U.S. Government, not subject to copyright.}
\affiliation{National Institute of Standards and Technology, Boulder,
  Colorado 80305, USA}
\author{M. J. \surname{Ablowitz}}
\affiliation{Department of Applied Mathematics, University of
 Colorado, Boulder, Colorado 80309-0526, USA}
\author{P. \surname{Engels}}
\affiliation{Department of Physics and Astronomy, Washington State
  University, Pullman, Washington 99164, USA}
\begin{abstract}
  The one-dimensional piston shock problem is a classical result of
  shock wave theory.  In this work, the analogous dispersive shock
  wave (DSW) problem for a dispersive fluid described by the nonlinear
  Schr\"{o}dinger equation is analyzed.  Asymptotic solutions are
  calculated using Whitham averaging theory for a "piston" (step
  potential) moving with uniform speed into a dispersive fluid at
  rest.  These asymptotic results agree quantitatively with numerical
  simulations.  It is shown that the behavior of these solutions is
  quite different from their classical counterparts.  In particular,
  the shock structure depends on the speed of the piston.  These
  results have direct application to Bose-Einstein condensates and the
  propagation of light through a nonlinear, defocusing medium.
\end{abstract}
\pacs{
  03.75.Kk, 
  03.75.Lm, 
  05.45.Yv, 
  47.40.Nm} 
\maketitle


The study of dispersive shock waves (DSWs) has gained interest with
the recent experimental realization of DSWs in a Bose-Einstein
condensate (BEC) \cite{Dutton2001,Hoefer2006} and the propagation of
light through a nonlinear, defocusing medium \cite{Wan2007}.
Comparisons between classical, viscous shock waves (VSWs) and DSWs
have been discussed in the context of single shocks \cite{Hoefer2006}
and the interaction of two shocks \cite{Hoefer2007} yielding some
appealing similarities but also important differences.  Motivated by
the classical VSW piston problem, here we consider the generation of a
DSW by a piston moving into a dispersive fluid at rest.

The theoretical study of DSWs involves averaging a periodic wave over
its period and allowing for slow variation of the wave's parameters.
This method, known as Whitham averaging \cite{Whitham1965}, has been
successfully applied to many DSW problems including step initial data
for the nonlinear Schr\"{o}dinger (NLS) equation
\cite{Gurevich1987,El1995}, Bose-Einstein condensates
\cite{Kamchatnov2004,Hoefer2006}, fiber optics \cite{Kodama1999}, the
generation of ultrashort lasers \cite{Biondini2006}, and DSW
interactions \cite{Hoefer2007}.  We also note that a dispersive piston
shock problem arises as an asymptotic reduction of 2D, supersonic flow
of a dispersive fluid around an obstacle \cite{El2004}.

The piston shock problem is one of the canonical problems in the
theory of VSWs.  A uniform gas is held at rest in a long, cylindrical
chamber with a piston at one end.  When the piston is impulsively
moved into the gas with constant speed, a region of higher density
builds up between the piston and a shock front which propagates ahead
of it.  An elegant asymptotic solution to this problem is well known
and relates the shock speed to the speed of the piston and the initial
density of the gas (see e.g.\ \cite{Courant1948} and the discussion at
the end of this work).

In this work, we consider the problem of a ``piston'' moving with
constant speed into a steady, dispersive fluid: e.\ g.\ a
Bose-Einstein condensate or light propagating through a nonlinear,
defocusing medium.  The piston in this case is a step potential that
moves with uniform velocity.  This potential could be realized in a
BEC with a repulsive dipole beam and in nonlinear optics with a local
change in the index of refraction.  One expects, in analogy with the
classical, viscous case, the generation of a dispersive shock wave.
As we will show, this is indeed the case.  There are two types of
asymptotic behavior depending on the piston velocity.  For smaller
piston velocities, a region of larger density or intensity builds up
between the piston and a DSW.  However, for large enough piston
velocities, a locally periodic wave is generated between the piston
and the DSW which has no VSW correlate.  The asymptotic results are
verified by numerical simulations demonstrating quantitative
agreement.


We consider the 1D NLS equation with a potential (also known as the
Gross-Pitaevskii (GP) equation)
\begin{equation}
  \label{eq:1}
  i \eps \Psi_{t} = - \frac{\eps^2}{2}
  \Psi_{xx} + V_0(x,t) \Psi + |\Psi|^2 \Psi, \quad 0 < \eps \ll 1.
\end{equation}
This equation models the mean field of a quasi-1D BEC
\cite{Perez-Garcia1998} and the slowly varying envelope of the
electromagnetic field propagating through a Kerr medium
\cite{Boyd2003} (where time $t$ is replaced by the propagation
distance).  The small parameter $\eps$ is inversely proportional to
the number of atoms in the BEC \cite{Hoefer2006} or, after rescaling,
inversely proportional to the maximum initial intensity of the
electromagnetic field.  For all calculations in this work, we assume
$\eps = 0.015$, a typical experimental value for BEC
\cite{Hoefer2006}.  The piston problem is modeled with a temporally
and spatially varying step potential given by
\begin{equation*}
  \begin{split}
    V_0(x,t) &= V_{max} H(v_pt-x), \quad H(y) = \left \{
      \begin{array}{cc}
        0 & y < 0 \\
        1 & y \ge 0
      \end{array} \right .,
  \end{split}
\end{equation*}
with strength $V_{max}$ and constant velocity $v_p$.  The initial
conditions are
\begin{equation*}
  \Psi(x,0) \to \sqrt{\rho_R} ~ \textrm{as}
    ~ x \to \infty, \quad \Psi(x,0) \to 0 ~ \textrm{as} ~ x \to -\infty.
\end{equation*}
Because the strength of the piston is large, $V_{max} \gg \rho_R$, the
density/intensity rapidly decay to zero near the origin.  We assume
that the wavefunction $\Psi$ is in the ground state of the step
potential $V_{max}H(-x)$ when $t \le 0$.  For all calculations in this
work, $\rho_R = 0.133$.

It is useful to view eq.\ \eqref{eq:1} in its hydrodynamic form by
making the transformation $\Psi = \sqrt{\rho} e^{\frac{i}{\eps}
  \int_0^x u(x',t) \, \ud x'}$ and inserting this expression into the
first two local conservation equations for the GP equation
\begin{equation}
  \label{eq:4}
  \begin{split}
    \rho_t + (\rho u)_x &= 0 \\[2mm]
    (\rho u)_t + \left( \rho u^2 + \tfrac{1}{2} \rho^2 \right)_x
    &= \frac{\eps^2}{4} (\rho (\log \rho)_{xx} )_x - \rho V_{0_x},
  \end{split}
\end{equation}
where $\rho$ is the dispersive fluid ``density'' and $u$ is the
dispersive fluid ``velocity''.  These equations are similar to the
Navier-Stokes (NS) equations of fluid dynamics except that the viscous
term of NS has been replaced by the dispersive term with coefficient
$\eps^2/4$.

Because the dispersive term in eq.\ \eqref{eq:4} is small, one expects
the generation of small wavelength $O(\eps)$ oscillations near a steep
gradient in the fluid variables.  Witham's method is to average an
exact periodic solution over fast oscillations and assume that the
wave's parameters (amplitude, frequency, wavelength, etc.) vary slowly
\cite{Whitham1965}.

We convert the piston DSW problem into a moving boundary value problem
where appropriate boundary conditions are imposed at the piston front.
First we solve the piston DSW problem assuming sufficiently small,
positive piston velocities $v_p$.  ``Small'' will be defined below.

We assume the piston strength is large for $x < v_p t$, so there is
negligible density there.  Assuming there is a jump from zero density
to the nonzero value $\rho_L$ with a fluid velocity $u_L$ we integrate
the first conservation law in eqs.\ \eqref{eq:4} across the jump to
find
\begin{equation}
  \label{eq:3}
  -v_p \rho_L + \rho_L u_L = 0 .
\end{equation}
This gives the first boundary condition at the piston
\begin{equation}
  \label{eq:6}
  u(v_pt, t) \equiv u_L = v_p ,
\end{equation}
the fluid velocity at the piston equals the piston velocity.  We
require a boundary condition for the density.

The theory of DSWs involves a system of quasi-linear, first order,
hyperbolic equations known as the Whitham modulation equations.  The
Whitham equations describe the slow evolution of a periodic wave's
parameters and must be solved in order to find the asymptotic DSW
solution.  The simplest, non-trivial solutions to these equations are
known as \emph{simple waves} where only one dependent variable is
varying in space and time, and the rest are constant.  In analogy with
gas dynamics, we assume a simple wave solution, but in this case to
the Whitham equations.  This determines a density $\rho_L$ at the
piston.  In order to connect to the uniform state ahead of the piston
$\rho_R < \rho_L$, we must have a single DSW for $v_p$ sufficiently
small ($v_p < 2\sqrt{\rho_R}$).  As we will show below, a ``vacuum
state'' is created when $v_p \ge 2\sqrt{\rho_R}$, and we find a
uniform traveling wave (TW) with speed $v_p$, instead of the constant
density $\rho_L$, adjacent to the DSW.  Later we verify with numerical
simulations that these assumptions are reasonable.  Now we derive the
asymptotic piston DSW.

At the time $t=0^+$, we assume that there is a discontinuity in the
fluid variables due to the impulsive motion of the piston at $t=0$
\begin{equation}
  \label{eq:8}
  \rho(x,0^+) = \left\{
    \begin{array}{cc}
      \rho_L & x = 0 \\
      \rho_R & x > 0
    \end{array} \right., ~
  u(x,0^+) = \left\{
    \begin{array}{cc}
      u_L=v_p & x = 0 \\
      u_R=0 & x > 0
    \end{array} \right. .
\end{equation}
This discontinuity is regularized by a slowly modulated, traveling
wave, periodic solution to eq. \eqref{eq:4} with $V(x,t) \equiv 0$
\cite{Gurevich1987}
\begin{equation}
  \label{eq:7}
  \begin{split}
    \rho(x,t,\theta) &= \lambda_3-[\lambda_3-\lambda_1] \text{dn}^2
    (\theta;m), ~ m = \frac{\lambda_2-\lambda_1}{\lambda_3 -
      \lambda_1} \\
    u(x,t,\theta) &= V - \sigma
    \frac{\sqrt{\lambda_1\lambda_2\lambda_3}}{\rho(x,t,\theta)},
    ~\sigma = \pm 1, ~ 0 < \lambda_1 < \lambda_2 < \lambda_3
    \\
    \pd{\theta}{x} &= \sqrt{\lambda_3-\lambda_1}/\eps, ~
    \pd{\theta}{t} = -V\sqrt{\lambda_3-\lambda_1}/\eps, \\
  \end{split}
\end{equation}
where the parameters $\lambda_i$, $i=1,2,3$ and $V$ satisfy
\begin{equation}
  \label{eq:17}
  \begin{split} 
    \lambda_1 = &\tfrac{1}{16}(r_1-r_2-r_3+r_4)^2, ~
    \lambda_2 = \tfrac{1}{16}(-r_1+r_2-r_3+r_4)^2, \\
    \lambda_3 = &\tfrac{1}{16}(-r_1-r_2+r_3+r_4)^2 , ~ V =
    \tfrac{1}{4}(r_1+r_2+r_3+r_4),
  \end{split}
\end{equation}
and $r_i = r_i(x,t)$ evolve according to the Whitham equations
\begin{equation*}
  \begin{split}
    \pd{r_i}{t} + v_i(r_1,r_2,r_3,r_4) \pd{r_i}{x} = 0 , \quad
    i=1,2,3,4. 
  \end{split}
\end{equation*}
The velocities $v_i$ are expressed in terms of complete elliptic
integrals of the first and second kind \cite{Gurevich1987}.

In order to find a simple wave solution to the Whitham equations, we
require that only one of the parameters $r_i$ spatially varies and
that the initial data for all the parameters $r_i$ properly
characterizes the initial data in eq.\ \eqref{eq:8} with the spatial
average of eq.\ \eqref{eq:7}.  We use the method of initial data
regularization \cite{Kodama1999,Biondini2006,Hoefer2006,Hoefer2007} to
find
\begin{equation}
  \label{eq:9}
  \begin{split}
    r_1 &\equiv -2\sqrt{\rho_R}, ~ r_2 \equiv
    2\sqrt{\rho_R}, ~ 
    r_4 \equiv 2v_p + 2\sqrt{\rho_R}, \\
    &r_3(x,0^+) = \left \{
      \begin{array}{cc}
        2\sqrt{\rho_R} & x = 0 \\
        2v_p + 2\sqrt{\rho_R} & x > 0
      \end{array} \right. , ~\sigma \equiv 1,\\
    &\rho(v_p t, t) \equiv \rho_L =
    (\frac{1}{2}v_p + \sqrt{\rho_R})^2 .
  \end{split}
\end{equation}
The last equation, the boundary condition for the density at the
piston, comes from the simple wave assumption.  Equations \eqref{eq:9}
give rise to a self similar solution for $r_3$ satisfying the implicit
relation
\begin{equation}
  \label{eq:10}
  v_3(r_1,r_2,r_3(x,t),r_4) = x/t.
\end{equation}
Using a nonlinear root finder, eq.\ \eqref{eq:10} is solved
numerically for each $x$ and $t$.  The values for $r_i$, $i=1,2,3,4$
are inserted into eqs.\ \eqref{eq:17} and \eqref{eq:7} to determine
the asymptotic DSW solution.  A pure DSW propagates ahead of the
piston with trailing and leading edge speeds respectively
\cite{Gurevich1987,Hoefer2006}
\begin{equation}
  \label{eq:11}
  v_s^- = \frac{1}{2}v_p + \sqrt{\rho_R}, ~ v_s^+ = \frac{2 v_p^2
    + 4 v_p \sqrt{\rho_R} + \rho_R}{v_p + \sqrt{\rho_R}} .
\end{equation}
\begin{figure}
  \centering
  \begin{tabular}{cc}
    \begin{minipage}{0.48\columnwidth}
      \includegraphics[width=\columnwidth]{
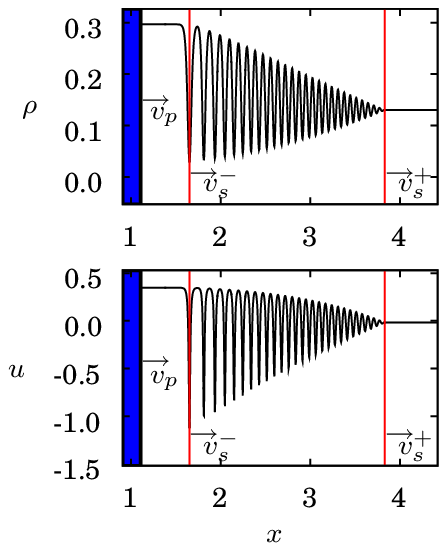}
    \end{minipage}
    &
    \begin{minipage}{0.48\columnwidth}
      \includegraphics[width=\columnwidth]{
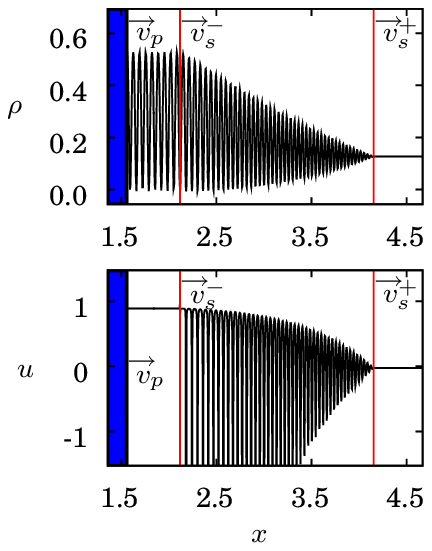}
    \end{minipage}
  \end{tabular}
  \caption{LEFT: Asymptotic piston DSW solution at time $t = 3$ for
    $v_p = \sqrt{\rho_R} = 0.365$.  The vertical lines mark the left
    and right edges of the DSW moving with speeds $v_s^- = 0.548$ and
    $v_s^+ = 1.278$ respectively.  RIGHT: Asymptotic piston DSW
    solution at time $t=1.7$ when $v_p = 2.5\sqrt{\rho_R} = 0.912$.  A
    locally periodic region connects the piston to the trailing edge
    of the DSW.  The density minima in this region are approximately
    zero and the velocity is, theoretically, undefined (infinite) at
    these points.  The density maxima are $4\rho_R = 0.532$.
    DSW speeds are $v_s^- = 1.245$, $v_s^+ = 2.452$. }
  \label{fig:piston_dsw}
\end{figure}
Figure \ref{fig:piston_dsw}, left depicts the asymptotic piston DSW
solution for a small piston velocity.  The minimum values of the
density and velocity occur at the trailing edge of the DSW and are
\cite{Gurevich1987,Hoefer2006}
\begin{equation}
  \label{eq:16}
  \rho_{min} = (\sqrt{\rho_R}-\frac{1}{2}v_p)^2, ~ u_{min} =
  -v_p(\frac{\sqrt{\rho_R} + \frac{1}{2} v_p}{\sqrt{\rho_R} -
    \frac{1}{2}v_p} ) .
\end{equation}
The maximum values occur between the piston and the DSW: $\rho_{max} =
\rho_L = (v_p/2 + \sqrt{\rho_R})^2$, $u_{max} = u_L = v_p$ .

\begin{figure}
  \centering
  \begin{tabular}{cc}
    \begin{minipage}{0.48\columnwidth}
      \includegraphics[width=\columnwidth]{
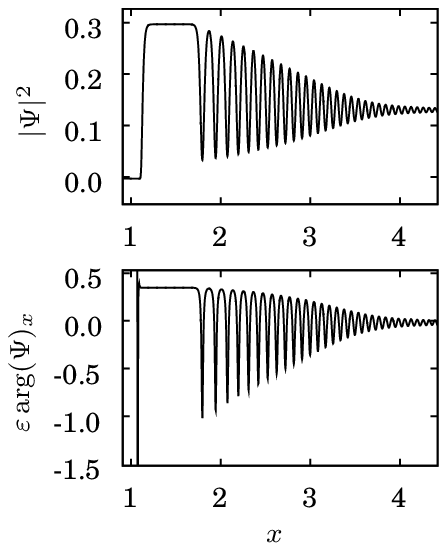}
    \end{minipage}
    &
    \begin{minipage}{0.48\columnwidth}
      \includegraphics[width=\columnwidth]{
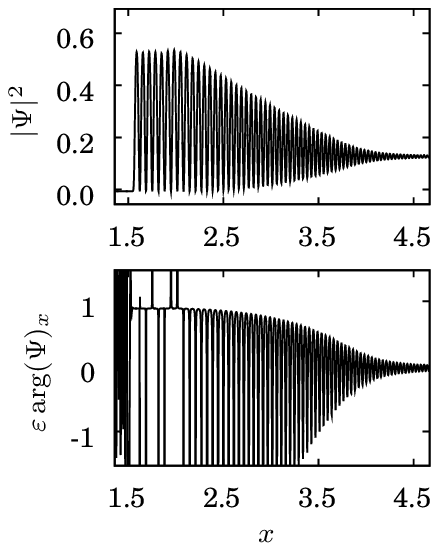}
    \end{minipage}
  \end{tabular}
  \caption{Numerical solutions to eq.\ \eqref{eq:1} for the piston
    problem.  LEFT: The density (upper) and velocity (lower) for the
    same parameters as in Fig.\ \ref{fig:piston_dsw} LEFT.
    Numerically calculated trailing edge speed is $0.557$,
    approximately the theoretical value $0.548$.  RIGHT: The density
    (upper) and velocity (lower) for the same parameters as in Fig.\
    \ref{fig:piston_dsw} RIGHT.  Numerically calculated TW velocity of
    locally periodic wave is $0.912$, approximately the piston speed
    $0.913$.}
  \label{fig:piston_dsw_numerics}
\end{figure}
It is possible for the piston velocity to be greater than the trailing
DSW velocity calculated using eq.\ \eqref{eq:11}
\begin{equation*}
  v_p \ge v_s^- \quad \text{if} \quad v_p \ge 2\sqrt{\rho_R} .
\end{equation*}
When $v_p = 2\sqrt{\rho_R}$, $\rho$ can vanish (there is a so-called
vacuum point) and a modification of the solution is required.  To find
a simple wave solution for large piston velocities, we must derive new
conditions for the parameters $r_i$.  We modify the DSW solution by
introducing a locally periodic TW between the piston and the trailing
edge of the DSW.  When $v_p = 2\sqrt{\rho_R}$, the DSW forms a vacuum
point \cite{El1995,Hoefer2006} at the piston.  The \emph{vacuum
  condition}, $\rho = 0$, is satisfied in eq.\ \eqref{eq:7} when
\begin{equation}
  \label{eq:13}
  \lambda_1 = 0 ~ \Rightarrow ~ r_1-r_2-r_3+r_4 = 0 ~ \Rightarrow ~
  r_4 - r_3 = 4\sqrt{\rho_R}.
\end{equation}
Note that the fluid velocity is undefined at a vacuum point, even
though the vacuum points have a well-defined propagation speed through
the fluid.  We assume that this condition holds for $v_p >
2\sqrt{\rho_R}$ as well.  One more condition is required to completely
determine $r_4$ and $r_3(x=0,t=0^+)$; $r_1$, $r_2$, and $r_3(x>0,0^+)$
are determined by the initial data \eqref{eq:8}.  Because there is a
locally periodic TW between the piston and DSW, we assume that the
velocity of the TW equals the piston velocity. This is the \emph{TW
  velocity condition}
\begin{equation}
  \label{eq:14}
  V = v_p ~ \Rightarrow ~ r_1+r_2+r_3 + r_4 = 4 v_p ~ \Rightarrow ~
  r_3 + r_4 = 4 v_p.
\end{equation}
Given the initial data in eq.\ \eqref{eq:8} and the two conditions
\eqref{eq:13} and \eqref{eq:14}, only $r_3$ and $\sigma$ in the
initial data of eq.\ \eqref{eq:9} are altered
\begin{equation*}
  r_3(x,0^+) = 2v_p \mp 2\sqrt{\rho_R}, ~
  \sigma(x,0^+) = \mp 1, ~ x \genfrac{}{}{0pt}{2}{=}{>} 0 .
\end{equation*}
Note that $\sigma = -1$ in the locally periodic region.  Figure
\ref{fig:piston_dsw}, right depicts the asymptotic DSW solution for
$v_p > 2\sqrt{\rho_R}$.

Several properties of this DSW solution are worth noting.  The density
between the piston and the DSW oscillates between the values
\begin{equation}
  \label{eq:21}
  \rho_{min} = 0 \quad \text{and} \quad \rho_{max} =
  4\rho_R,
\end{equation}
\emph{independent} of the piston velocity $v_p$ and the TW in this
region propagates with the velocity $V = v_p$.  Note that since the
vacuum condition \eqref{eq:13} holds everywhere inside the TW trailing
the DSW, the velocity in this region, from eq.\ \eqref{eq:7}, is $u =
V = v_p$ everywhere (except at the vacuum points where the velocity is
undefined).  The wavelength of the TW is $l = \frac{2 \eps}{v_p}
K(4\rho_R/v_p^2)$, where $K(m)$ and $E(m)$ are the complete elliptic
integrals of the first and second kinds respectively.  The DSW
propagates with trailing edge speed (also the propagation speed of the
rightmost vacuum point where $\sigma$ changes sign)
\begin{equation*}
  v_s^- = v_p + (v_p+3\sqrt{\rho_R})\left[ \frac{v_p
      E(4\rho_R/v_p^2)}{(v_p-2\sqrt{\rho_R}) K(4\rho_R/v_p^2)} - 1
  \right]^{-1}, 
\end{equation*}
and leading edge speed $v_s^+$, the same as that given in eq.\
\eqref{eq:11}.  The number of vacuum points increases linearly with
time:
\begin{equation}
  \label{eq:20}
  N_{vac}(t) \approx \left \lceil \frac{v_s^- - v_p}{l} t \right \rceil =
  \left \lceil \frac{(v_s^- - v_p)v_p}{2\eps 
      K(4\rho_R/v_p^2)} t  \right \rceil .
\end{equation}

\begin{figure}
  \centering
  \begin{tabular}{cc}
    \begin{minipage}{0.48\columnwidth}
      \includegraphics[width=\columnwidth]{
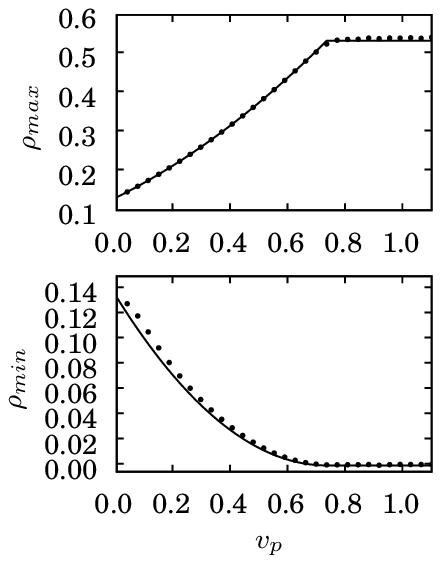}
    \end{minipage}
    &
    \begin{minipage}{0.48\columnwidth}
      \includegraphics[width=\columnwidth]{
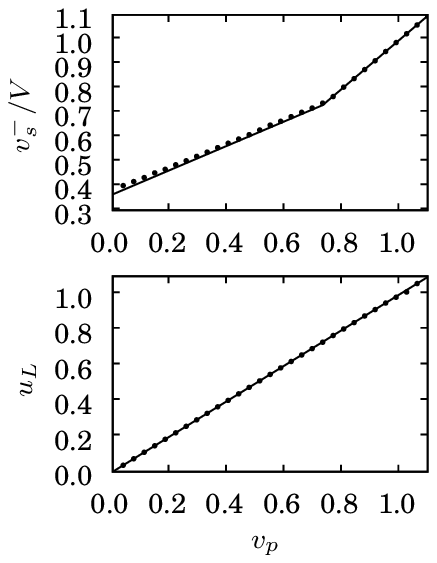}
    \end{minipage}
  \end{tabular}
  \caption{Comparisons of the analytical theory (solid curves) and
    numerical simulations (dots) for different DSW parameters as
    functions of the piston velocity $v_p$.  LEFT: $\rho_{max}$ (top)
    and $\rho_{min}$ (bottom).  Maximum absolute error in $\rho_{max}$
    from eqs.\ \eqref{eq:9} and \eqref{eq:21} is 0.0088.  Maximum
    absolute error in $\rho_{min}$ from eq.\ \eqref{eq:16} is 0.010.
    RIGHT: Top depicts the speed of the trailing edge of the piston
    DSW for $v_p < 2\sqrt{\rho_R} = 0.73$ and TW velocity $V$ of
    locally periodic wave for $v_p \ge 2\sqrt{\rho_R}$.  Bottom shows
    the validity of the boundary condition \eqref{eq:6}. Maximum
    absolute error in $v_s^-/V$ from eqs.\ \eqref{eq:11} and
    \eqref{eq:14} is 0.018.  Maximum absolute error in $u_L$ from eq.\
    \eqref{eq:6} is 0.013.}
  \label{fig:piston_dsw_comparisons}
\end{figure}

We perform direct numerical simulations of eq.\ \eqref{eq:1} to verify
the assumptions we have made such as the boundary conditions
\eqref{eq:6}, \eqref{eq:9}, the vacuum and TW velocity
conditions \eqref{eq:13}, \eqref{eq:14}, and the trailing edge DSW
speed $v_s^-$ of eq.\ \eqref{eq:11}.  All of our assumptions are in
excellent agreement with numerical simulation as shown in Fig.\
\ref{fig:piston_dsw_comparisons}.  
Numerical simulations of eq.\ \eqref{eq:1} were performed with the
pseudo-spectral, Fourier method \cite{Weideman1986} with a grid
spacing $\Delta x = 0.004$, time step $\Delta t = 0.0006$, parameter
$V_{max} = 5$, and a slightly smoothed step potential $V_0$.  The
initial data is relaxed in the presence of the potential $V_0$ and
spatially localized by including a smoothed step potential with
strength $V_{max}$ near the right boundary of the spatial domain.

Numerically calculated piston DSWs for both moderate $v_p =
\sqrt{\rho_R}$ and large $v_p = 2.5 \sqrt{\rho_R}$ piston velocities
are shown in Fig.\ \ref{fig:piston_dsw_numerics}.  For the slower
piston velocity in Fig.\ \ref{fig:piston_dsw_numerics} left, the
solution is similar to the asymptotic result in Fig.\
\ref{fig:piston_dsw} left.  The only difference is the variation in
the density and velocity through the piston.  In the model problem
considered here, we assume that the density goes to zero immediately
behind the piston.  The piston DSW corresponding to the large piston
velocity in Fig.\ \ref{fig:piston_dsw_numerics} right is very similar
to the asymptotic result in Fig.\ \ref{fig:piston_dsw} right.  The
vacuum condition in eq.\ \eqref{eq:13} predicts $u = v_p$ everywhere
in the trailing wave region except at vacuum points where it is
undefined.  This is reflected in the numerical calculation as very
large spikes in the velocity when the density approaches zero.

The analogous piston viscous shock wave problem in shallow water is
discussed in, e.g.\ \cite{Courant1948}; the 1d equations are
equivalent to eqs.\ \eqref{eq:4} when $\eps = 0$, $V \equiv 0$ and a
dissipative regularization is used whenever a shock forms.  The
asymptotic solution is found by assuming a simple wave and
incorporating the boundary condition \eqref{eq:6}.  In this case, one
finds that the shock speed ($v_s$) is always larger than the piston
speed ($v_p$), i.e.\ one finds $v_s - v_p = v_p \rho_R/(\rho_L-\rho_R)
> 0$.

The analysis in this work shows that techniques from VSW theory,
simple wave solutions and suitable jump conditions, are useful in the
study of DSWs.  Nevertheless, DSWs can lead to very different
phenomena.

\acknowledgments{This research was partially supported by the U.S. Air
  Force Office of Scientific Research, under grant FA4955-06-1-0237;
  by the National Science Foundation, under grant \mbox{DMS-0602151};
  and by the National Research Council.}


\bibliographystyle{apsrev}

\end{document}